\documentclass[aps,prx,groupedaddress,superscriptaddress,twocolumn,showpacs,longbibliography]{revtex4-1}

\usepackage{graphicx}
\usepackage{dcolumn}
\usepackage{bm}
\usepackage[T1]{fontenc}
\usepackage[french]{babel}
\usepackage{epstopdf}
\usepackage{amssymb}
\usepackage{color}
\usepackage{amsmath}
\usepackage{subfigure}
\usepackage{natbib}

\newcommand{\be}{\begin{equation}}
\newcommand{\ee}{\end{equation}}
\newcommand{\bea}{\begin{eqnarray}}
\newcommand{\eea}{\end{eqnarray}}

\begin{document}

\preprint{APS/123-QED}

\title{Origin and magnitude of `designer' spin-orbit interaction in graphene on semiconducting transition metal dichalcogenides}

\author{Zhe Wang}
\author{Dong-Keun Ki}
\affiliation{Department of Quantum Matter Physics (DQMP) and Group of Applied Physics (GAP), University
of Geneva, 24 quai Ernest-Ansermet, CH-1211 Gen\`{e}ve 4, Switzerland}

\author{Jun Yong Khoo}
\affiliation{Department of Physics, Massachusetts Institute of Technology, Cambridge, Massachusetts 02139, United States}

\author{Diego Mauro}
\affiliation{Department of Quantum Matter Physics (DQMP) and Group of Applied Physics (GAP), University
of Geneva, 24 quai Ernest-Ansermet, CH-1211 Gen\`{e}ve 4, Switzerland}

\author{Helmuth Berger}
\affiliation{Institut de Physique de la Mati\`{e}re Complexe, Ecole Polytechnique Federale de Lausanne, CH-1015 Lausanne, Switzerland}

\author{Leonid S. Levitov}
\affiliation{Department of Physics, Massachusetts Institute of Technology, Cambridge, Massachusetts 02139, United States}

\author{Alberto F. Morpurgo}
\affiliation{Department of Quantum Matter Physics (DQMP) and Group of Applied Physics (GAP), University
of Geneva, 24 quai Ernest-Ansermet, CH-1211 Gen\`{e}ve 4, Switzerland}


\begin{abstract}
We use a combination of experimental techniques to demonstrate a general occurrence of spin-orbit interaction (SOI) in graphene on transition metal dichalcogenide (TMD) substrates. Our measurements indicate that SOI is ultra-strong and extremely robust, despite it being merely interfacially-induced, with neither graphene nor the TMD substrates changing their structure. This is found to be the case irrespective of the TMD material used, of the transport regime, of the carrier type in the graphene band, and of the thickness of the graphene multilayer. Specifically, we perform weak antilocalization (WAL) measurements as the simplest and most general diagnostic of SOI, and show that the spin relaxation time is  very short (approximately 0.2 ps or less) in all cases regardless of the elastic scattering time, whose value varies over nearly two orders of magnitude. Such a short spin-relaxation time strongly suggests that the SOI originates from a modification of graphene band structure. We confirmed this expectation by measuring a gate-dependent beating, and a corresponding frequency splitting, in the low-field Shubnikov-de Haas magneto-resistance oscillations in high quality bilayer graphene devices on WSe$_2$. These measurements provide an unambiguous diagnostic of a SOI-induced splitting in the electronic band structure, and their analysis allows us to determine the SOI coupling constants for the Rashba term and the so-called spin-valley coupling term, i.e., the terms that were recently predicted theoretically for interface-induced SOI in graphene. The magnitude of the SOI splitting is found to be on the order of 10 meV,  more than 100 times greater than the SOI intrinsic to graphene. Both the band character of the interfacially induced SOI, as well as its robustness and large magnitude make graphene-on-TMD a promising system to realize and explore a variety of spin-dependent transport phenomena, such as, in particular, spin-Hall and valley-Hall topological insulating states.

\end{abstract}

\maketitle
\section{Introduction}
Van der Waals (vdW) heterostructures formed by vertical stacks of different two-dimensional (2D) materials have emerged recently as designer systems, providing a new paradigm for engineering novel electronic media with widely tunable parameters~\cite{Geim2013}. Stacked vdW heterostructures nicely combine the ability to tailor interfacial interactions at the atomic scale, while at the same time preserving the integrity of individual layers. This `designer' aproach is epitomized by recent work on graphene paired with hexagonal boron-nitride (hBN)~\cite{Dean2013,Ponomarenko2013,Hunt2013}. In this system, a dramatic change in the graphene band structure occurs when the crystal axes of graphene and hBN layers are nearly aligned, in the total absence of any reorganization of chemical bonding or any change in the atomic order of individual layers. The transformed band structure manifests itself in striking and robust transport phenomena --such as the appearance of so-called satellite Dirac points-- that are readily observable experimentally. These unexpected findings are opening up a wide avenue of research exploring vdW heterostructures based on many different 2D materials ~\cite{Yu2013,Britnell2013,Lee2014,Li2015,Rivera2015,ZWang2015}. A key goal at this stage is to identify the interfacial interactions that can alter specific electronic properties of interest, and to  understand the microscopic physical processes responsible for their origin.

One fascinating question in this vein is whether vdW heterostructures can be used to control not only the orbital dynamics of electrons in graphene, but also their spin, i.e., whether vdW heterostructures can be employed to generate a strong spin-orbit interaction (SOI) in the graphene Dirac band. To this end, combining graphene with large-gap semiconducting transition metal dichalcogenides (TMDs) appears to be a promising route~\cite{Gmitra2015}, because semiconducting TMDs exhibit an extremely strong SOI ~\cite{Zhu2011,Xiao2012,Kosmider2013,Wang2012}, and because they are known to preserve the high electronic quality of graphene when used as substrates~\cite{Kretinin2014}. Recent magneto-transport measurements performed on graphene-on-WS$_2$ ~\cite{Avsar2014,Wang2015}-- and in particular the observation of a pronounced weak antilocalization (WAL) contribution to the conductivity of graphene~\cite{Wang2015}-- confirm these expectations. Indeed, the analysis of the experimental results indicated that the spin relaxation time $\tau_{so}$ in graphene-on-WS$_2$ is between 100 and 1000 times shorter than in graphene on SiO$_2$ ~\cite{Tombros2007} or on hBN~\cite{Guimaraes2014}. This is broadly consistent with {\it ab-initio} calculations, which predict the strength of SOI in graphene-on-WS$_2$ to be at least 100 times larger than the SOI intrinsic to graphene~\cite{Wang2015,Huertas2006,Min2006,Konschuh2010}.

These results pose a number of interesting and challenging questions, which are central for our understanding of the new phenomenon of `designer' SOI. In particular, the physical process by which strong SOI can be imprinted by one layer on an adjacent layer without any changes in their structure remains puzzling. So far, the experiments were unable to elucidate the microscopic mechanism responsible for the strong SOI in graphene, nor did they provide any reliable insight into the functional form of the induced SOI as well as its strength. They also did not establish whether a strong interfacially-induced SOI is unique to graphene-on-WS$_2$ or whether it is a robust, generic property of all graphene-on-TMD heterostructures. Last but not least, perhaps the most tantalizing question of all is whether SOI is dominated by disorder scattering or by a band structure modification. If the latter happens to be the case, the strong SOI present in graphene-on-TMD system can be employed to create and explore a variety of electronic media with novel properties~\cite{Kane2005,Kane20052,Gmitra2016}.

Here we exploit a variety of graphene-on-TMD heterostructures to tackle these questions in a comprehensive way. Large part of the work focuses on the study of WAL in heterostructures formed by graphene and one of the semiconducting TMDs: WSe$_2$, MoS$_2$, and WS$_2$. For all TMDs used, irrespective of carrier mobility ($\mu$), position of the Fermi level in graphene, and thickness of the graphene layer (up to trilayer), a pronounced WAL signal is observed. This finding shows that interfacially induced SOI in graphene-on-TMDs is an extremely robust phenomenon, insensitive to virtually all details of the vdW heterostructure considered. A quantitative analysis of the WAL data allows us to establish an upper bound of approximately 0.2 ps for the spin-relaxation time $\tau_{\rm so}$, irrespective of the carrier mobility (which was varied by nearly two orders of magnitude). Such a short $\tau_{\rm so}$ value appears to be physically compatible only with SOI originating from a modification of the graphene band structure. To validate this conclusion we present measurements of Shubnikov-de Haas (SdH) conductance oscillations exhibiting a beating due to a splitting in their frequency~\cite{Das1989,Nitta1997,Grbi2008}. The size of the splitting and its dependence on carrier density show that the beating originates from SOI, and its quantitative analysis allows us to establish the SOI magnitude. We find that the dominant SOI term is of the Rashba type, and that its characteristic energy is approximately $\lambda_R \simeq 10-15$ meV; the strength of the other SOI term expected to be induced by interfacial interactions~\cite{Wang2015} --i.e., the one that couples spin and valley-- ranges between $\lambda=0$  and $\lambda\approx$ 5-6 meV (i.e., experimental data are compatible with $\lambda_R = 15$ meV and $\lambda = 0$ meV or with $\lambda_R = 10$ meV and $\lambda = 5-6$ meV, as well as different choices in these intervals).

Besides elucidating most aspects of interfacially induced SOI in graphene, the results presented here clearly illustrate the experimental flexibility of graphene-on-TMDs heterostrcutures. These heterostructures allow comparative studies by varying the specific TMD material used, the thickness of the graphene layer, the position of the Fermi level, and the scattering time ($\tau$). Even more flexibility could be introduced using double gated devices, to tune the graphene band structure (e.g., in bilayers)~\cite{Oostinga2008,Weitz2010,Velasco2012,Craciun2009} and further extend the range of carrier densities accessible experimentally. We anticipate that this unrivaled experimental flexibility will prove useful in future experiments aiming at exploring other aspects of interfacial interactions in vdW heterostructures.

\begin{figure*}[t]
\begin{center}
\includegraphics[width=0.8\textwidth]{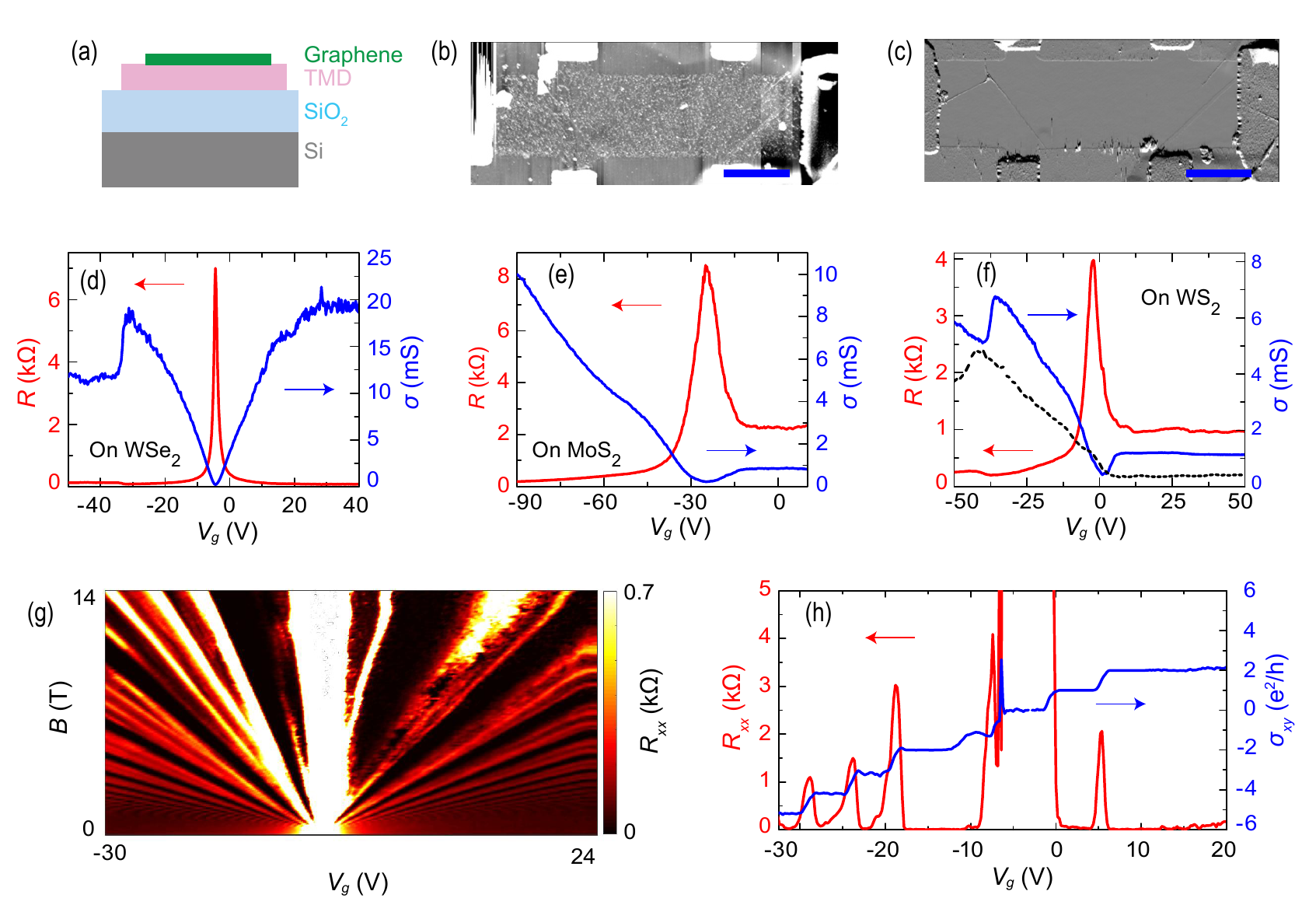}
\caption{(Color online) Basic characterization of graphene on TMD substrates. (a) Schematic cross section of the devices. The graphene layer (green) is transferred onto a TMD crystal (pink) that has been previously exfoliated on a substrate consisting of highly doped Silicon (gray) covered by 285-nm-thick SiO$_2$ (light blue). The silicon substrate is operated as a back-gate. (b-c) AFM images of a graphene Hall-bar device before and after the AFM-ironing process (scale bar is 2 $\mu$m long). (d-f) Gate-voltage ($V_g$) dependence of the resistance ($R$; red curves) and the conductivity ($\sigma$; blue curves) of monolayer graphene on WSe$_2$ (d), MoS$_2$ (e) and WS$_2$ (f), measured at 4.2 K. The carrier mobility in the three cases is 110,000, 33,000, and 23,400 cm$^2$/Vs, respectively. In all devices, the conductance saturates at large enough $V_g$ away from charge neutrality point, when charges start to be accumulated at the SiO$_2$-TMD interface. The black line in (f) represents the $V_g$-dependent conductivity $\sigma(V_g)$ before AFM-ironing (the corresponding mobility is approximately two times smaller than for the blue curve). (g-h) Integer quantum-Hall effect (QHE) observed in high-quality graphene-on-WSe$_2$ at $T=250$ mK whose basic characterization is shown in (d). The color map of the longitudinal resistance ($R_{xx}$) versus $V_g$ and $B$ (g) and the $V_g$-dependence of $R_{xx}$ (red curves) and the Hall conductance ($\sigma_{xy}$; blue curves) measured at $B=12$ T (h) clearly confirm the occurrence of the vanishing $R_{xx}$ and concomitantly quantized $\sigma_{xy}=\nu \times e^2/h$ at integer values of filling factor ($\nu \equiv nh/eB$). In panel (h), the QH plateaus at $\nu=\pm1$, -3, -4, -5 due to the full degeneracy lifting of the $N=0$, 1 Landau levels in monolayer are clearly visible. }
\end{center}
\label{FIG1}
\end{figure*}

\section{Device fabrication and characterization}
Van der Waals heterostructures were assembled by transferring graphene layers of different thickness (mono, bi, or tri layers) onto thin exfoliated flakes of TMDs (WSe$_2$, WS$_2$, and MoS$_2$), resulting in devices whose cross section is schematically shown in Fig. 1(a). For the assembly, we employed a commonly used dry transfer technique~\cite{Dean2010,Nuno2014}. Conventional electron-beam lithography, lift-off, and oxygen plasma etching techniques were employed to pattern and contact multi-terminal Hall-bar devices. The contacts consist of an evaporated Ti/Au thin film (10/70 nm). All structures were realized on substrates consisting of degenerately doped Silicon covered with a 285 nm thick layer of thermally grown SiO$_2$. The charge density ($n$) of graphene is tuned by operating the doped Silicon substrate as a gate electrode. In this configuration, for sufficiently large gate voltage ($V_g$), carriers are accumulated at the surface of the TMD flake at the interface with the SiO$_2$ layer~\cite{Wang2015,Kim2015}. When that happens, the carrier density in graphene cannot be tuned anymore by $V_g$, and the conductivity ($\sigma$) of graphene saturates.

The device quality --i.e., the carrier mobility and the inhomogeneity in carrier density-- depends on the details of the assembly process, and on the procedure to ``clean'' the device structure at the end of the fabrication. Two key elements introduce disorder: structural defects resulting from the transfer process, such as ``bubbles'' and ``wrinkles'' in the graphene layer~\cite{Mayorov2011}, and adsorbates adhering onto graphene (mostly polymer residues remaining at the end of the fabrication process). The influence of both elements can be controlled in different ways. Structural defects can be eliminated by defining the graphene Hall-bar in regions in which these defects are absent. Selecting these areas, which typically have linear dimensions of the order of 5-6 $\mu$m, usually results in very high carrier mobility: we have observed low-temperature mobility values as large as 160,000 cm$^2$/Vs, comparable to (or possibly even slightly better than) the best values  observed in graphene-on-hBN structures assembled by the same dry-transfer technique~\cite{Dean2013,Ponomarenko2013,Hunt2013,Young2012,Taychatanapat2013}. Selecting larger areas is also possible, but this unavoidably prevents the full exclusion of structural defects, resulting in lower mobility.

Adsorbates can be eliminated in a rather controlled way by a so-called atomic force microscope (AFM) ``ironing'' process~\cite{Goossens2012}, essential to realize high mobility devices. AFM ironing consists of scanning the graphene flake with an AFM tip in contact mode, applying only a moderate force, in such a way to pile up all the adsorbates just outside the edges of the graphene flake. The effectiveness of the process is illustrated in Figs. 1(b-c). Fig. 1(b) shows an AFM image of a device at the end of fabrication and Fig. 1(c) another image of the same device taken after the ``ironing'' step. The difference --the extremely small corrugation that is measured on graphene after the ``ironing'' process-- is clear. In our studies we have measured  16 different devices, in which --depending on their area, cleaning procedure adopted, density of ``bubbles'', etc.-- the carrier mobility extracted from measurements of the conductivity and of Hall effect ranged between 3,000 and 160,000 cm$^2$/Vs.

Figs. 1(d-f) shows the gate voltage dependence of the resistance (red curves) and of the corresponding conductivity (blue curves) measured on three representative devices, respectively on WSe$_2$, MoS$_2$, and WS$_2$. As compared to our earlier work on graphene-on-WS$_2$, in which no AFM ironing was done~\cite{Wang2015}, in the current generation of higher-quality devices the charge neutrality point is ``exposed'' in all cases: it is possible to shift the Fermi level ($E_F$) both in the valence and in the conduction band by acting on the back gate. Whereas for graphene on WS$_2$ and MoS$_2$ only a small range of energies in the conduction band can be accessed, for WSe$_2$, $E_F$ can be shifted over a rather large interval in both the valence and the conduction band. Hence, WSe$_2$ allows the systematic investigation of SOI in the conduction band without the need to use a top gate electrode, something that could not be done in previous work. Finally, Figs. 1(g-h) show that all integer quantum Hall effect (QHE) states are visible, including the symmetry broken states caused by the presence of electron-electron interactions~\cite{Young2012}, which is indicative of the high quality of the devices (in the best cases, symmetry broken states become already visible for applied magnetic fields as low as approximately 1 Tesla).

\begin{figure*}[t]
\begin{center}
\includegraphics[width=0.8\textwidth]{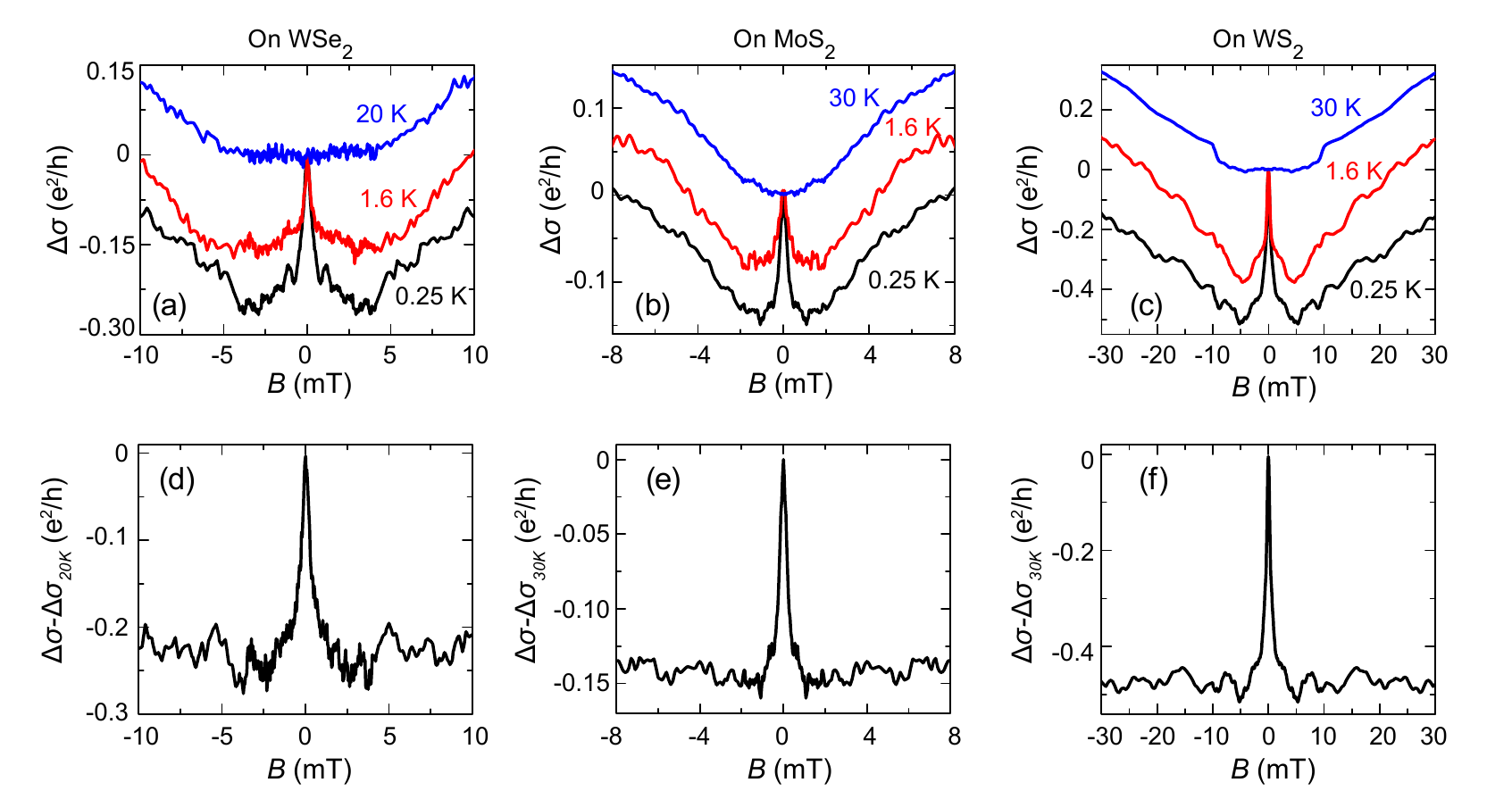}
\caption{(Color online) Negative magneto-conductivity due to WAL in monolayer graphene on TMD substrates. (a-c) Ensemble-averaged magneto-conductivity measured on monolayer graphene on WSe$_2$, MoS$_2$ and WS$_2$, for $T=250$ mK (black curve), $T=1.6$ K (red curves), and $T=20 or 30$ K (blue curves), with the Fermi level gate-tuned to be in the graphene valence band. The data are measured on the same devices whose $V_g$-dependent transport curves are shown Figs. 1(d-f). The characteristic peak due to WAL around $B=0$ T is clearly visible at low temperature and disappears at higher temperature. (d-f) Quantum corrections to magneto-conductivity of monolayer graphene on WSe$_2$, MoS$_2$, and WS$_2$, obtained by subtracting the classical contribution (corresponding to the magneto-conductivity measured at $T=20$ or 30 K) from the magneto-conductivity measured at $T=250$ mK. Note that up to the highest magnetic field investigated, no signatures of weak-localization are visible. }
\end{center}
\label{FIG2}
\end{figure*}

\section{Extracting SOI from weak antilocalization in monolayer graphene on different TMDs}

Weak antilocalization is a striking quantum interference effect originating from spin-orbit coupling that has long served as a direct probe of SOI  in conductors~\cite{Hikami1980,Beenakker1991}. Also for graphene-on-TMD heterostructures, the observation of the WAL correction to the low-temperature magneto-conductivity provides the simplest and most general diagnostic of the presence of SOI ~\cite{Wang2015}. Extracting the WAL contribution requires suppressing the effect of the so-called phase-coherent universal conductance fluctuations (UCF) originating from random interference of electronic waves~\cite{Lee1985}. Indeed, since the dimensions of our graphene-on-TMD devices are typically comparable to (or even smaller than) the phase coherence length $L_{\phi}$, the WAL contribution in any individual measurement is normally eclipsed by the presence of UCF. To make the WAL contribution stand out, we suppress the magnitude of the UCF by looking at the ensemble averaged conductivity, obtained by averaging many (typically 50) magneto-conductance traces measured at slightly different values of $V_g$ (the procedure is identical to that described in Ref.~\cite{Wang2015} to which we refer for details).

The procedure described above, performed at different temperatures, leads to the results shown in Figs. 2(a-c) for graphene on WSe$_2$, MoS$_2$, and WS$_2$ respectively. In all cases, a negative magneto-conductivity of order $e^2/h$ is clearly apparent at the lowest temperature investigated, $T=250$ mK, upon the application of a magnetic field ($B$) of a few milliTesla. The magnitude of the negative magneto-conductivity decreases upon warming up the devices and the effect disappears entirely at $T\simeq 20-30$ K, as expected for quantum interference effects~\cite{Beenakker1991}. We conclude that, irrespective of the TMD used to realize the heterostrcutures, the presence of a pronounced WAL signal in magneto-transport demonstrates that in all cases SOI is induced in graphene. 

\begin{figure*}[t]
\begin{center}
\includegraphics[width=0.8\textwidth]{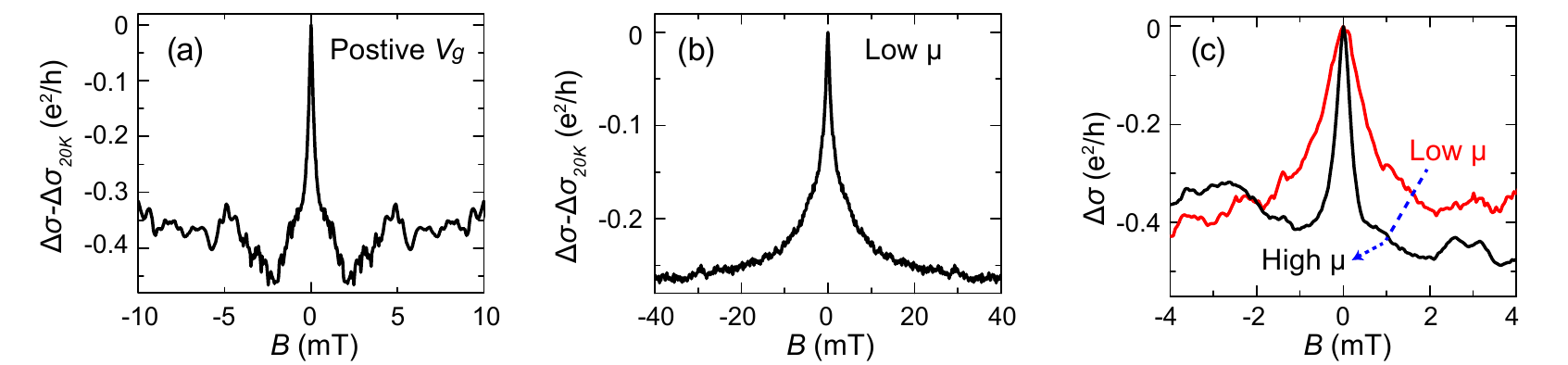}
\caption{(Color online) Interfacially induced SOI in graphene-on-TMD is a robust phenomenon. (a) Magneto-conductivity of graphene-on-WSe$_2$ measured with the Fermi energy gate-tuned to be in the graphene conduction band. (b) Magneto-conductivity due to WAL measured on a larger-area graphene-on-WSe$_2$ device with a carrier mobility of only 3,000 cm$^2$/Vs. (c) Magneto-conductivity of graphene-on-WS$_2$ measured on a same device before (red) and after (black) the AFM-ironing process needed to clear graphene from adsorbates. As the carrier mobility increases, the WAL peak becomes sharper as expected. All data in this figure have been measured at $T=250$ mK.}
\end{center}
\label{FIG.3}
\end{figure*}

In extracting the strength of SOI from magnetotransport measurements a special care should be taken in accounting for the interplay between WAL and weak localization, since the two effects contribute to magnetoconductivity with opposite signs~\cite{Bergmann1983}. Weak localization gives rise to a positive magneto-conductivity, which becomes visible when during the phase-coherent progapation of electrons their spins do not rotate-or rotate by a small enough angle~\cite{Hikami1980}. Despite the presence of SOI, a small positive magneto-conductivity due to weak localization may still be observed, because at sufficiently high magnetic field only the shortest trajectories give a non-negligible contribution to the interference effects probed by the ensemble-averaged conductivity. Unless SOI is extremely strong, the electron spin may not have time to rotate by a sufficient amount along these short trajectories, and  signatures of weak localization may then become visible at large $B$.

The issue is relevant because --as it is clear from Figs. 2(a-c)-- a positive magneto-conductivity is visible in the measurements, and it is important to establish whether this is a manifestation of weak localization. To this end, we recall that WAL and weak localization are quantum corrections to the conductivity, i.e. they correspond to the difference between the total magneto-conductivity that is actually measured and the classical contribution. The classical contribution is straightforward to determine, since it corresponds to the magneto-conductivity measured at sufficiently high temperatures, where phase coherent effects have been suppressed because of the thermally induced shortening of $L_{\phi}$. Since at 20-30 K the effect of WAL has entirely disappeared and the remaining low-field (positive) magneto-conductivity does not exhibit any significant temperature dependence, we can take the magneto-conductivity measured at these temperatures to be a good approximation of the classical contribution. The resulting quantum correction to the magneto-conductivity for the different devices is shown in Figs. 2(d-f). Within the precision of the measurements, determined by the remnant amplitude of UCF fluctuations, no positive magneto-conductivity is visible in Figs. 2(d-f). We conclude that irrespective of the TMD material used in the heterostructure, a clear WAL signal is always present with no detectable weak localization contribution. This observation provides a first clear indication that in all heterostructures investigated, the SOI induced in graphene has a very strong intensity.

Data measured on other monolayer graphene devices confirm that WAL always occurs with no detectable weak localization signal, irrespective of whether carriers are electrons or holes, and of their mobility (or, equivalently, scattering time $\tau$), which we varied over a range of nearly two orders of magnitude. Both aspects had not been addressed in our previous work on WS$_2$, in which the Fermi level could not be shifted into the conduction band and the scattering time was only varied by a limited amount~\cite{Wang2015}. The occurrence of WAL for electron transport is best illustrated with data measured on graphene-on-WSe$_2$, shown in Fig. 3(a), in which a fully developed WAL signal is clearly visible. The effect of the mobility can be appreciated by looking at Figs. 3(b) and 3(c). Fig. 3(b) shows the quantum correction to the magneto-conductivity measured on a large-area graphene-on-WSe$_2$ device, in which a high-density of ``bubbles''  led to a low-temperature mobility of 3,000 cm$^2$/Vs (whereas all devices shown in Figs. 2(a-c) had mobility larger than 25,000 cm$^2$/Vs). We find that the magnitude of the WAL correction is comparable in all cases, but in the lower mobility devices the magnetic field required to observe the negative magneto-conductivity is larger. This is expected, since when the mobility is lower a larger magnetic field is needed to pierce a flux of $\Phi_0=h/e$ through the area in which the electronic waves propagate phase coherently and interfere. The same conclusions can be drawn by looking at Fig. 3(c), which shows the WAL magneto-conductivity in a same graphene-on-WS$_2$ device measured before (red curve) and after (black curve) performing an AFM ``ironing'' step, resulting in a mobility increase. It is apparent that in this case as well, a higher mobility leads to a decrease of the magnetic field scale needed to suppress the effect of WAL.

\begin{figure*}[t]
\begin{center}
\includegraphics[width=0.8\textwidth]{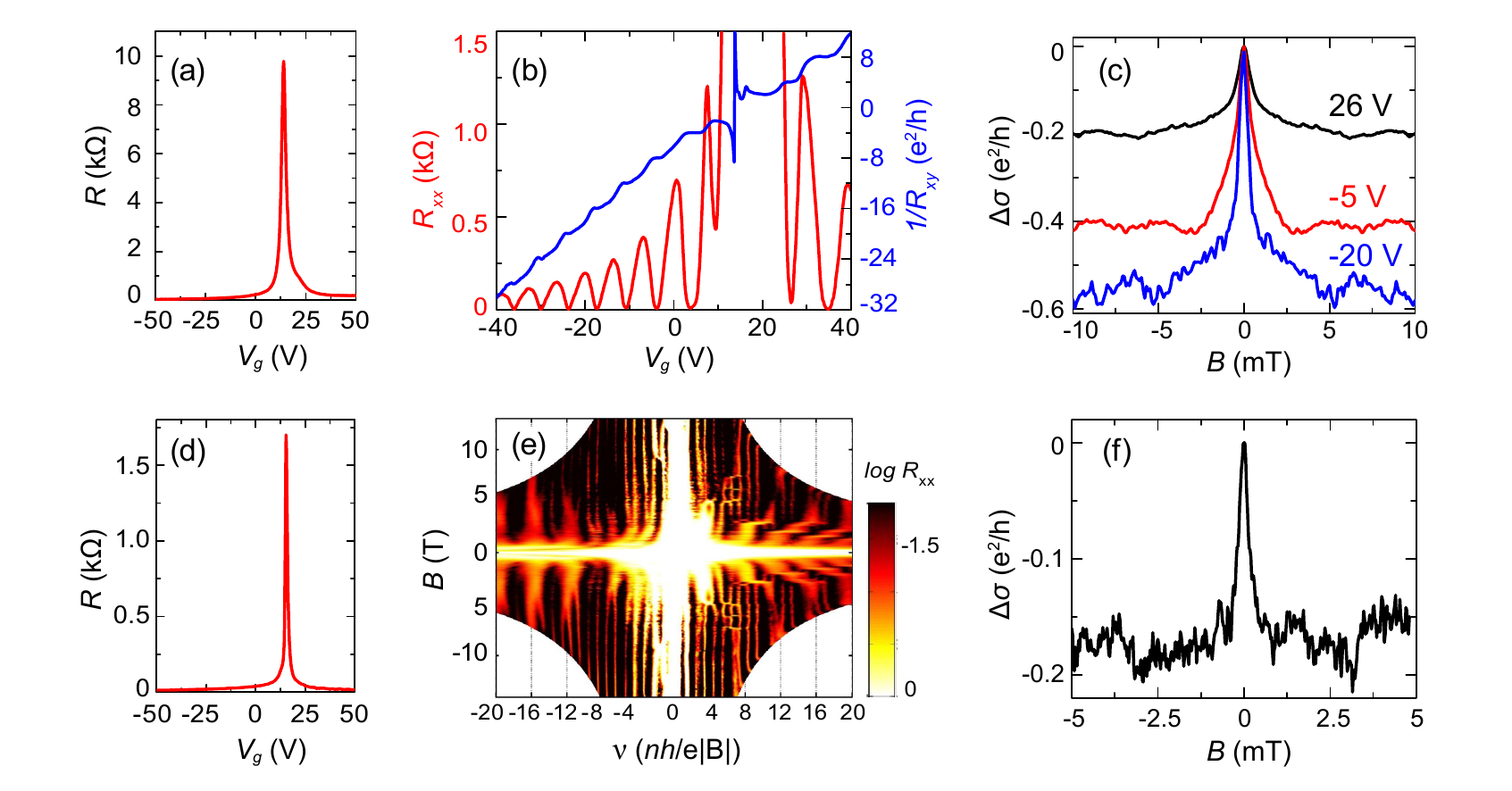}
\caption{(Color online) Interfacially-induced SOI in graphene multilayers on TMDs. Panels (a-c) and (d-f) refer respectively to a bilayer and a trilayer device, both realized on WSe$_2$. Panels (a) and (d) show that the resistance peak around charge neutrality is very narrow, indicative of a high uniformity of carrier density (in the trilayer, the magnitude of inhomogeneity is smaller than 2$\times$10$^{10}$ cm$^{-2}$, comparable to the best devices on hBN). High quality integer quantum Hall effect is observed in both devices, as shown by data in panels (b) and (e) measured at $T=4.2$ K. In panel (b), in which the characteristic quantization sequence of graphene bilayer is observed, the applied magnetic field is $B=4$ T. Panel (e) shows a color plot of the longitudinal resistance of trilayer graphene as a function of filling factor and magnetic field. Full breaking of the degeneracy of Landau levels due to electron-electron interactions start to be clearly visible already at $B=1$ T. Panels (c) and (f) show the ensemble averaged magneto-conductivity measured at $T=250$ mK, exhibiting the characteristic signature of WAL, with no positive magneto-conductivity due to weak localization (in panel (c) the different curves show data at different gate voltages, corresponding to shifting the Fermi level from the bilayer graphene valence band to the conduction band).}
\end{center}
\label{FIG.4}
\end{figure*}

\section{SOI in bilayer graphene and thicker multi-layers}
When monolayer graphene is substituted by thicker multilayers, a strong SOI remains present in all devices (irrespective of the TMD used for the device realization). Fig. 4 illustrates this conclusion with data measured on two different graphene-on-WSe$_2$ devices: Figs. 4(a-c) refer to a bilayer graphene device with low-temperature mobility $\mu = 33,000$ cm$^2$/Vs and Figs. 4(d-f) to a trilayer having mobility $\mu \simeq 110,000$ cm$^2$/Vs. Basic transport characterization show that --as for monolayer devices-- heterostructures based on thicker multilayers exhibit an excellent electronic quality. The resistance peak around the charge neutrality point is extremely sharp in both cases (see Figs. 4(a) and 4(d)); in the trilayer device the measured width corresponds to a charge inhomogeneity as low as 1.8$\times$10$^{10}$ cm$^{-2}$, comparable to the best ever reported for non-suspended graphene devices. In the presence of a perpendicular magnetic field $B=4$ T the expected Hall effect quantization sequence is observed in the bilayer device, with plateaus in the Hall conductance occurring at $\sigma_{xy}=4Ne^2/h$ ($N= \pm 1, \pm 2, ...$)~\cite{Novoselov2006,McCann20062}, concomitantly with the vanishing of the longitudinal resistance. In the thicker multilayer, the plot of the longitudinal resistance versus filling factor $\nu \equiv nh/eB$ and $B$ (Fig. 4(e)) shows the appearance of broken symmetry quantum Hall states already at $B$ as low as approximately 1 T. In short, excellent quality bi and tri-layer graphene devices can be realized on TMD substrates, comparable to the very best devices realized on hBN by means of the same technique.

Figs. 4(c) and 4(f) show that a pronounced low temperature negative magneto-conductivity due to WAL is clearly visible in both the bilayer and the trilayer device. In these devices as well, no background due to weak localization is observed, indicative of the large SOI strength. This is remarkable, because interfacial interactions are expected to modify only the properties of the bottom graphene layer, the one in direct contact with the TMD crystal~\cite{Wang2015}. Under normal conditions, the eigen-functions in the different bands of the multilayer are such that electrons have a finite amplitude of probability to be in that layer. As a result, all bands are affected and that is why thicker multilayers exhibit a pronounced WAL. Nevertheless, it is clear that the effect of interfacially induced SOI should decrease in intensity as the thickness of the multilayer increases, since the probability for electrons to enter in contact with the TMD crystal decreases upon increasing thickness (or, equivalently, the amplitude of the electron wavefunctions in the bottom layer --the one in contact with the the TMD-- decreases for thicker multilayers). The data shown in Fig. 4(c) --and especially those shown in Fig. 4(f)-- indicates that despite the larger thickness, at least up to trilayer graphene the observed behavior of the WAL correction is the one typical of very strong SOI. Significantly thicker multilayers are needed to ``dilute'' the effect of SOI induced by interfacial interactions.

The observation of WAL in graphene bilayers (Fig. 4(c)) is worth an additional comment. In monolayers, WAL can occur due only to the Dirac nature of electrons (i.e., in the absence of SOI), as a consequence of the $\pi$ Berry phase picked up by the electron wave functions that undergo backscattering processes while staying in a same valley. WAL due to this effect has been seen experimentally, albeit only at elevated temperatures (typically $T \geq 10$ K)~\cite{Tikhonenko2009}, since only then $L_{\phi}$ is sufficiently short (such a dependence on temperature allow the phenomenon to be discriminated from WAL due to SOI, which increases in amplitude upon cooling). The effect is absent in bilayer graphene, since in bilayers a $2\pi$ Berry phase is acquired by the electron wavefunction upon back-scattering, which does not lead to WAL~\cite{Gorbachev2007}. As such, the occurrence of WAL in bilayers illustrated by the data shown in Fig. 4(c) provides unambiguous and more direct evidence of the presence of interfacially induced SOI.

In concluding this Section, we emphasize that the possibility of using interfacial interactions with a TMD substrate to induce strong SOI in different graphene multilayers --and not only in monolayers-- adds to  the flexibility of this experimental system. As we will discuss in Section VI, we exploit this flexibility already in the present work to determine quantitatively the type and magnitude of the interfacially induced SOI.

\begin{figure}[t]
\begin{center}
\includegraphics[width=1\linewidth]{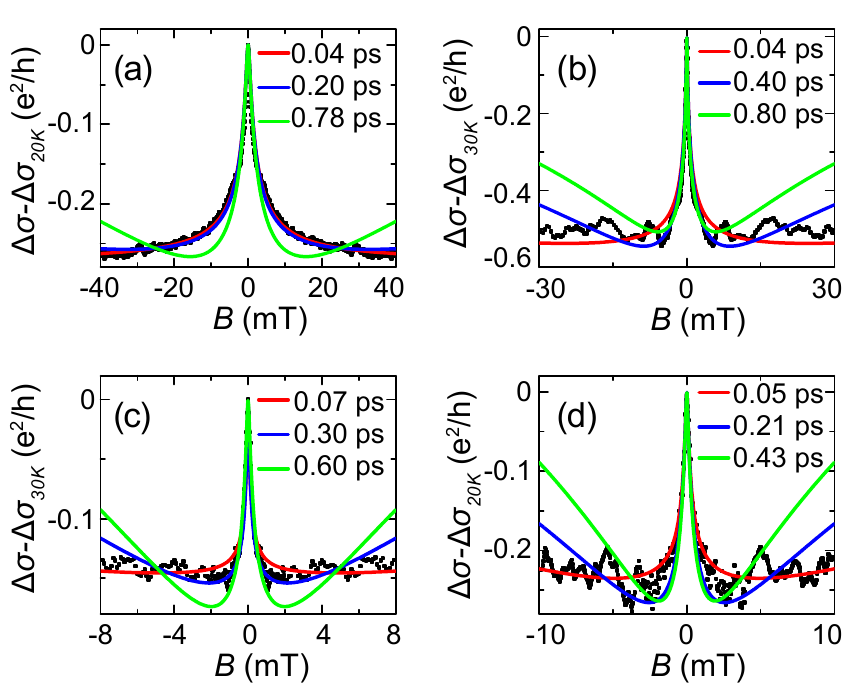}
\caption{(Color online) Comparison of WAL data measured on monolayer graphene on different TMDs with the theoretical predictions of Eq. (1). In panels (a-d), the black dots represent the data; the continuous lines of different colors represent the predictions of Eq. (1) for different values of spin-relaxation time $\tau_{\rm so}$. In panel (a), data from a low-mobility device ($\mu = 3,000$ cm$^2$/Vs) on WSe$_2$ are shown; Panels (b-d) show data on higher mobility devices (respectively, 23,400, 33,000, and 110,000 cm$^2$/Vs) on WS$_2$ (b), MoS$_2$ (c), and WSe$_2$ (d). Note in all cases, inserting values of $\tau_{\rm so}>0.5$ ps in Eq. (1) leads to the appearance of a positive magneto-conductivity due to weak localization at higher $B$, which is not seen in the experiments. This allows us to determine an upper bound for $\tau_{\rm so}$ for all the devices investigated. All data in this figure have been measured at $T=250$ mK.}
\end{center}
\label{FIG.5}
\end{figure}

\begin{figure}[t]
\begin{center}
\includegraphics[width=0.8\linewidth]{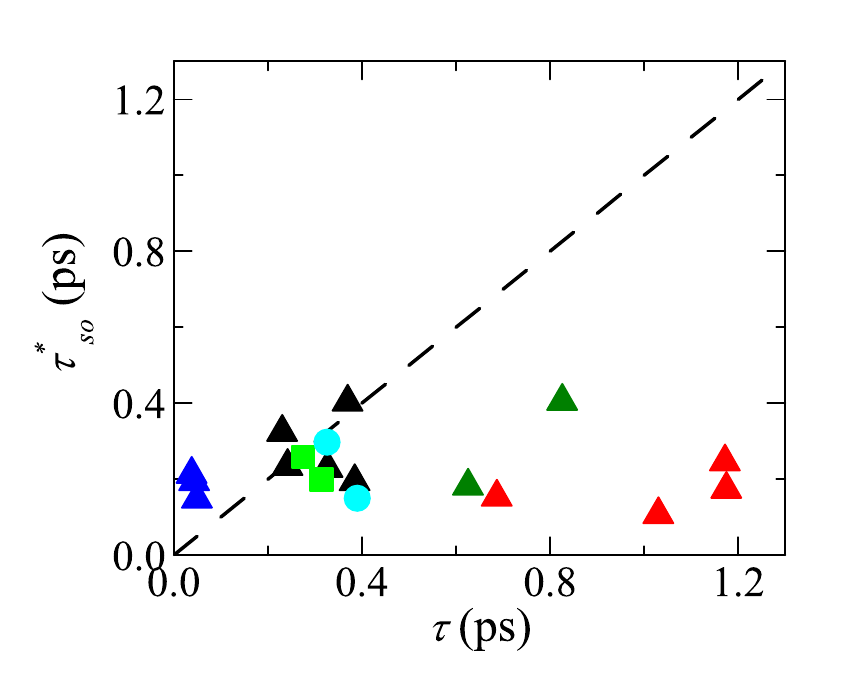}
\caption{(Color online) Upper bound of $\tau_{\rm so}$, extracted from WAL data, as a function of $\tau$ for monolayer graphene devices. The shape of the different symbols refers to devices realized on  different TMD substrates (triangles for WSe$_2$, circles for MoS$_2$, and squares for WS$_2$). For each different symbol, data points represented with the same color refer to the same device measured at different gate voltages. The broken line corresponds to $\tau_{\rm so}=\tau$.}
\end{center}
\label{FIG.6}
\end{figure}

\section{Quantitative analysis of the weak antilocalization data}
The pronounced WAL observed in the measurements presented in the previous sections, together with the absence of any signature of weak localization, illustrate that strong SOI is a general property of graphene-TMD heterostructures. For a quantitative analysis, we confine ourselves to devices realized on monolayers, for which an explicit theoretical expression for the magneto-conductivity due to WAL is available~\cite{McCann2012}. The expression, which takes into account the effect of all possible symmetry-allowed SOI terms, reads:
\begin{eqnarray}\nonumber
\Delta \sigma(B)& =& -\frac{e^2}{\pi h}\left[F\left(\frac{\tau^{-1}_B}{\tau^{-1}_\phi}\right) -F\left(\frac{\tau^{-1}_B}{\tau^{-1}_\phi + 2\tau^{-1}_{\rm asy}}\right)\right. \\
 && \left. - 2F \left(\frac{\tau^{-1}_B}{\tau^{-1}_\phi + \tau^{-1}_{\rm so}}\right)\right].
\label{WALeq}
\end{eqnarray}
where $F(x)=\ln(x)+\psi(1/2+1/x)$ with $\psi(x)$ the digamma function.
Here $\tau_B^{-1}=4DeB/\hbar$ ($D$ is the carrier diffusion constant), $\tau_\phi^{-1}$ is the dephasing rate, $\tau_{\rm asy}^{-1}$ represents the spin relaxation rate due to the SOI terms that break $z\rightarrow-z$ symmetry ($z$ is the direction normal to the graphene plane) and $\tau_{\rm so}^{-1}$ is the total spin-relaxation rate, including the effect of all SOI terms.

Since weak (anti)localization theory is developed having the fully diffusive transport regime in mind~\cite{McCann2012}, we start with the analysis of the magneto-conductivity measured in the lowest mobility device, i.e. the graphene-on-WSe$_2$ device whose data are shown in Fig. 3(b) (carrier mobility $\mu \cong 3,000$ cm$^2$/Vs; elastic scattering time $\tau \approx 0.04$ ps). Fig. 5(a) shows that Eq. (1) reproduces the data well with $\tau_{\rm so} \approx 0.2$ ps, with all other parameters satisfying the conditions of validity of the theory: $\tau_\phi > \tau_{\rm asy} > \tau_{\rm so} > \tau$~\cite{Note1}. Some more considerations are however needed to understand physically the meaning of the good agreement between Eq. (1) and the experimental data.

To this end, we note that, for $\tau_{\rm so}$ values longer than 0.2 ps, theory predicts that weak localization should become visible in the magnetic field range explored in the measurements (see, for instance, the green curve in Fig. 5(a), which represents Eq. (1) with $\tau_{\rm so} = 0.78$ ps), contrary to what is observed experimentally (no positive magneto-conductivity is observed up to the largest magnetic field $B=40$ mT applied in the experiments). $\tau_{\rm so}$ values significantly larger than 0.2 ps are therefore not compatible with our observations and can be excluded. $\tau_{\rm so}$ values shorter than 0.2 ps, however, are compatible with the measurements. This is illustrated by the red curve in Fig. 5(a), which is a plot of Eq. (1) with $\tau_{\rm so} = 0.04$ ps (and with all other parameters to be the same). This choice for $\tau_{\rm so}$ also perfectly reproduces the data and leads to a magneto-conductivity that is indistinguishable from the case $\tau_{\rm so} = 0.2$ ps. We therefore conclude that our analysis of WAL can only provide an upper value for $\tau_{\rm so} \approx 0.2$ ps.

These considerations make it clear that measuring WAL up to sufficiently high magnetic field is important, because the absence of a positive magneto-conductivity due to weak localization is what allows a more precise quantitative determination of the upper bound for $\tau_{\rm so}$. In our earlier work on graphene-on-WS$_2$~\cite{Wang2015}, in which the analysis of WAL was also used to obtain $\tau_{\rm so}$, the magnetic field range had not been extended sufficiently in the measurements. As a result, the estimates of $\tau_{\rm so}$ reported there are approximately one order of magnitude larger than what we find now. In other words, from the analysis reported in our previous work, the intensity of SOI --albeit already very strong-- appeared to be weaker than what it actually is.

For devices with higher mobility, in which the electron mean free path $l=v_F \tau$ becomes longer, the analysis of WAL requires more critical thinking. For instance, the use of Eq. (1) is meaningful only up to magnetic field values $B \approx \Phi_0/l^2$ (where $\Phi_0=h/e$ is the quantum of flux), because in the diffusive regime described by Eq. (1) the minimum area of a time reversed trajectory giving rise to WAL or weak localization is of the order of $l^2$~\cite{Bergmann1983}. For devices in which $\mu \simeq 100,000$ cm$^2$/Vs, the corresponding value of $B$ is only approximately 5 mT. This does not pose problems to observe the characteristic peak in WAL, which for such high $\mu$ values becomes extremely narrow (see, e.g., Fig. 5(c)), but limits the range of magnetic field that can be meaningfully used in the quantitative analysis of the data. Nevertheless, in practice, we find that Eq. (1) does reproduce satisfactorily the measured magneto-conductivity in all cases, as illustrated in Figs. 5 (b-d). Therefore, we proceed as we discussed for the low mobility case, and determine the upper bound for $\tau_{\rm so}$ for each of the devices analyzed. This upper bound is such that for larger $\tau_{\rm so}$ values the predictions of Eq. (1) shows the presence of a positive magneto-conductivity at higher magnetic field and are incompatible with the experimental observations, whereas for smaller values of $\tau_{\rm so}$ the magneto-conductivity predicted by Eq. (1) does not change significantly and reproduces the behavior of the experimental data.

The result of this analysis is summarized in Fig. 6, in which the upper bound for $\tau_{\rm so}$ extracted for all monolayer devices analyzed is plotted as a function of the elastic scattering time $\tau$. For each device, the result of the analysis performed for different applied gate voltage is also shown. The upper bounds for $\tau_{\rm so}$ determined from WAL cluster between 0.1 and 0.4 ps, and in most cases they are close to 0.2 ps. This is a remarkably systematic behavior, especially considering the large range in carrier mobility investigated (from 3,000 to 110,000 cm$^2$/Vs). 

We note that care is needed in interpreting this result, because for sufficiently high mobility devices (e.g., in all cases in which $\tau > 0.5$ ps in Fig. 6) $\tau_{\rm so} < \tau$, which appears to be beyond the regime of validity of Eq. (1) (the assumption that motion is diffusive implies that $\tau$ is the shortest time scale). This situation is not new. It has already been encountered in the analysis of WAL in different two-dimensional systems in which SOI is known to be extremely strong, such as 2D hole gases in GaAs-heterostructures~\cite{Miller2003,Grbi2008}. As discussed in detail in Ref. ~\cite{Grbi2008}, also in these systems a pronounced signal due to WAL is observed without any positive magneto-conductivity due to weak localization. The quantitative analysis of the magneto-conductivity gives an upper limit for $\tau_{\rm so}$ ($\tau_{\rm so} \approx 3$ ps in that case), such that $\tau > \tau_{\rm so}$ ($\tau \approx 25$ ps for those systems), in complete analogy to what we find in our high mobility graphene-on-TMD devices. For 2D holes in GaAs-heterostructures, the very short $\tau_{\rm so}$ values extracted from the analysis of WAL were taken as a signature of a strong SOI originating from the band structure (i.e., not from impurity scattering)~\cite{Grbi2008}. This conclusion was validated through a  study of Shubnikov-de Haas (SdH) resistance oscillations, that exhibit a gate-voltage dependent beating. The beating is due to SOI that splits the hole Fermi surface, causing SdH oscillations to occur with two distinct frequencies. By analyzing the frequency splitting as a function of carrier density, the precise nature of the SOI term present in the Hamiltonian could be established.

In view of the similarity in the behavior of WAL in graphene-on-TMDs and in GaAs-based 2D hole gases it is tempting to draw analogous conclusions. Namely, the observed behavior of WAL appears to indicate a band origin of SOI in graphene-on-TMD. Indeed, the SOI-induced splitting in the electronic band structure can generate WAL which is at most weakly dependent on the amount of disorder in the system. This is consistent with our observation of WAL which is strong and robust for a wide range of carrier mobilities in a variety of different samples. To confirm the band origin of interfacially induced SOI, we now proceed to search for the occurrence of a beating in the low-field Shubnikov-de Haas (SdH) resistance oscillations.

\section{Spin-Orbit Band Structure Splitting and SdH oscillations in high-mobility graphene-on-TMD devices}

The SdH oscillations of transport coefficients in non-quantizing magnetic fields arise due to cyclotron motion of carrier states at the Fermi level. The periodic dependence on the inverse field $1/B$ provides a convenient way to measure the Fermi surface size. In the presence of spin-orbital splitting, the electronic band structure gives rise to split Fermi surfaces with different spin polarization. In this regime, the SdH oscillations exhibit a characteristic beating pattern that provides an unambiguous diagnostic of the split Fermi surface, allowing to directly measure the spin splitting value.

Measurement of the beating patterns in SdH oscillations relies on resolving a large number of Landau levels at moderate-to-low magnetic fields. Achieving this regime requires devices of exceptional quality. One constraint arises from carrier mobility which must be high enough to prevent the washing out of Landau levels of high order. Another constraint, which is equally important, is the absence of significant inhomogeneity in carrier density across the device. Indeed, an inhomogeneous density would result in washing out of the SdH oscillations due to different parts of the device contributing to the SdH oscillations with different frequencies. If the spread in frequencies originating from the carrier inhomogeneity is comparable to (or larger than) the SOI-induced frequency splitting, no splitting can be detected experimentally.

The high mobility values that can be achieved in our graphene-on-TMD devices are comfortably in the range needed for detecting a SOI-induced beating in the SdH oscillations. Charge inhomogeneity, on the contrary, poses certain challenges. Previous work indicates that, in the density range of our experiments, the inhomogeneity effects are less prominent in graphene bilayer (BLG) as compared to graphene monolayer~\cite{Efetov2016}. This is so because the density of quasiparticle states is higher in the bilayer, where quasiparticle dispersion is quadratic, and lower in the monolayer, where the dispersion is linear~\cite{Castro2009}. We therefore employ high-mobility BLG devices for this part of the experiments.

As shown in Fig. 7(a), magneto-resistance measurements performed on high-quality BLG-on-WSe$_2$ indeed exhibit beating in the SdH oscillations. The node of the beating pattern, marked by arrows, shifts towards higher magnetic field values when a more negative gate voltage is applied. Accordingly, the SdH oscillations Fourier spectrum (Fig. 7(b)) exhibits a pair of peaks with the splitting that increases upon shifting $V_g$ further away from charge neutrality. The full behavior is illustrated in Fig. 7(c) which shows the Fourier spectrum as a function of $V_g$ and oscillation frequency $f$. The frequencies corresponding to the maxima of the split peaks in the Fourier spectrum are plotted in Fig. 7(d) as a function of $V_g$. The observed splitting in the SdH oscillation frequency is a direct manifestation of the SOI splitting of the Fermi surface of BLG-on-WSe$_2$.

\begin{figure}[t!]
\begin{center}
\includegraphics[width=1\linewidth]{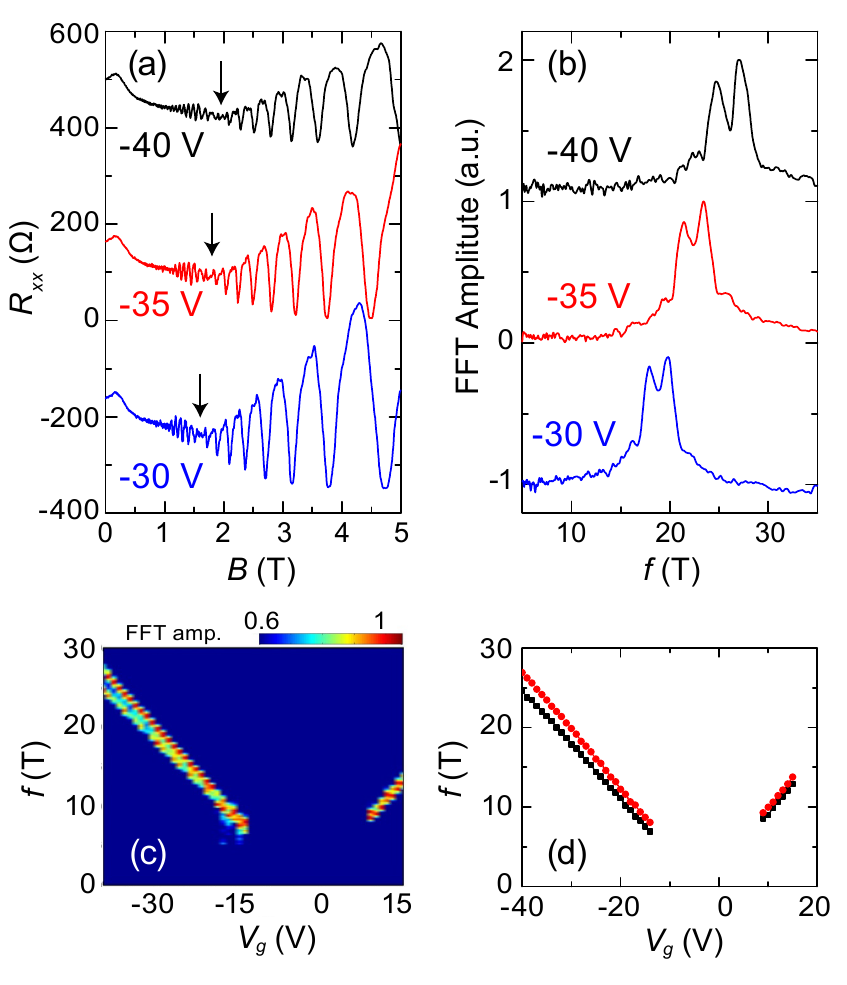}
\caption{(Color online) Extracting SOI from the SdH resistance oscillations observed in BLG-on-WSe$_2$ devices. (a) SdH resistance oscillations exhibit a beating pattern whose node position shifts with the applied gate voltage $V_g$ (curves are vertically offset for clarity; the charge neutrality point in this device is at $V_g=-2$ V). (b) Peak splitting in the Fourier spectra of the data shown in (a) is used to determine the SdH frequency splitting. (c) Color-coded Fourier spectrum plotted vs. frequency $f$ and gate voltage $V_g$. (d) The position of the two peaks in the Fourier spectra shown in (c) plotted vs. $V_g$ (black and red circles represent the lower and higher frequency peaks, respectively). The dependence on $V_g$ indicates that the SdH frequency splitting increases as the carrier density increases.}
\end{center}
\label{FIG.7}
\end{figure}

\begin{figure}[t!]
\begin{center}
\includegraphics[width=0.9\linewidth]{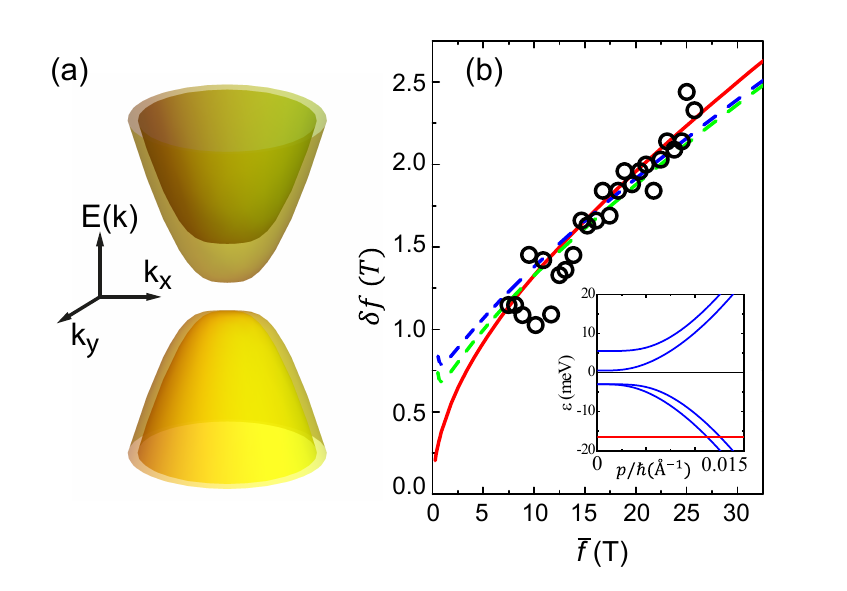}
\caption{(Color online) (a) Low energy band structure of bilayer graphene on TMD, obtained from the Hamiltonian in Eq. (3), with $U=-6$ meV, $\lambda_R=13$ meV, $\lambda=5$ meV. (b) The circles represent the measured splitting $\delta f$ of the frequency of the SdH resistance oscillations plotted versus the average frequency peak $\overline{f}$. The values of $\delta f$ and $\overline{f}$ are obtained from the data shown in Fig. 7(d) as the Fermi energy is swept through the valence band ($V_g<-2$ V). The colored lines correspond to theoretical predictions calculated using the band structure obtained with the Hamiltonian in Eq.(3) for different values of $\lambda_R$ and $\lambda$. Three possible best-fit plots with $(\lambda_R, \lambda)=(15, 0)$ meV (red solid line), $(\lambda_R, \lambda)=(13, -5)$ meV (green, dashed line), and $(\lambda_R, \lambda)=(13, 5)$ meV (blue, dashed-dotted line)) are shown. The inset shows cross section of Fig. 8(a), with the red line indicating the position of the Fermi energy calculated self-consistently (see text).}
\end{center}
\label{FIG.8}
\end{figure}

These findings indicate that the effect of the interfacially induced SOI is dominated by a modification of the band structure of graphene. For each value of $V_g$, the two peak frequencies are proportional to the areas of the two split Fermi surfaces. The measured dependence of these areas on $V_g$ can be used to reconstruct the band structure. The general Hamiltonian for BLG-on-TMD is of the form of a low-energy BLG Hamiltonian with an SOI term added to describe  interfacial coupling to TMD. Literature describes BLG in terms of two Dirac Hamiltonians of the two constituent monolayers coupled by the $\gamma_1$ term describing inter-layer nearest neighbor hopping~\cite{Castro2009,McCann20062,Edward2013}. The effect of interfacially induced SOI can be modeled by an effective Hamiltonian which was determined through \textit{ab-initio} calculations in a previous study of monolayer graphene on WS$_2$~\cite{Wang2015}. The expression for the spin-orbit Hamiltonian obtained in Ref.~\cite{Wang2015} is
\be \label{eq:H_soi}
H_{SOI} = \frac{\lambda}{2}\tau_zs_z1_\sigma+\frac{\lambda_R}{2}(\tau_z\sigma_xs_y-\sigma_ys_x),
\ee
where Pauli matrices $s_{x,y,z}$  and $\sigma_{x,y,z}$ represent electron spin-$1/2$ and pseudo-spin (sublattice A-B wavefunction components), respectively. The parameters $\lambda$ and $\lambda_R$ represent the strengths of the two types of SOI induced by TMD substrate, hereafter referred to as ``spin-valley SOI'' and Rashba SOI, respectively~\cite{Wang2015}\cite{Delta}.  

Here we incorporate the Hamiltonian given in Eq.\eqref{eq:H_soi} in the part of the full BLG-on-TMD Hamiltonian describing the graphene layer in direct contact with the WSe$_2$ substrate. We find the minimal single-valley Hamiltonian describing BLG-on-WSe$_2$ that reads:
\begin{eqnarray}\nonumber
H& =& v(\tau_z\sigma_xk_x+\sigma_yk_y)1_s1_\rho +\frac{1}{2}\gamma_11_s(\sigma_x\rho_x+\sigma_y\rho_y) \\
 && + \frac{1}{2}U1_s1_\sigma\rho_z+\left(\frac{1_\rho+\rho_z}{2}\right)H_{SOI},
\label{BLGHso}
\end{eqnarray}
where $\rho_{x,y,z}$ are Pauli matrices corresponding to the BLG layer index, the quantites $\sigma_{x,y,z}$ and $s_{x,y,z}$ were defined above, the quantities $1_{\rho,\sigma,s}$ denote the corresponding 2$\times$2 identity matrices, and $\tau_z=\pm1$ is the valley index (valley degeneracy persists in the presence of SOI).
The Hamiltonian $H$ includes the inter-layer potential difference $U$. In the single-gated configuration of our devices, the value $U$ is finite in the presence of an applied gate voltage $V_g$~\cite{Edward2013}.

We use the Hamiltonian given in Eq. (\ref{BLGHso}) for the quantitative estimate of the SOI parameters $\lambda$ and $\lambda_R$. This is done by determining the SdH oscillation frequency $f$ of each spin-split band from the area of the corresponding Fermi surface and comparing the resulting values with the experimental data. To this end, we employ the approach of Ref.~\cite{McCann2006} for obtaining $E_F$ self-consistently. For a given $\lambda$, $\lambda_R$ and $U$, the band structure can be easily computed, see Fig. 8(a) as an example. Due to the layer asymmetry, eigenstates of the Hamiltonian Eq. (\ref{BLGHso}) are not equally partitioned between the two graphene layers of the BLG. Thus, just as in the case of the spin-degenerate gapped BLG, the total carrier density $n = n_1 + n_2$ fixed by a given $E_F$ is unequally split between the two layers. The unequal layer carrier densities give rise to interlayer potential difference $\tilde{U}(n)$ that depends on the total carrier density $n$ \cite{sdhfitting}. By varying $E_F$ (or $n$), we can find the self-consistent value for which $\tilde{U}(n) = U$. The self-consistent solution corresponds to the experimental situation in which the gate voltage $V_g$ is given by $C_{\square} (V_g-V_{g0})=en$, where $V_{g0}$ corresponds to the charge neutrality point and $C_{\square}$ is the capacitance per unit area between the device and gate electrode. As an illustration, the red line in inset of Fig. 8(b)  indicates the position of the self-consistent value of $E_F$ for $U=-6$ meV, $\lambda_R=13$ meV, and $\lambda=5$ meV.

Knowing the self-consistent $E_F$ for a given $U$, we can then determine the Fermi momenta $k_{F+}$ and $k_{F-}$ of the two spin-split bands, or --equivalently-- the areas of the corresponding Fermi surfaces and the density of carriers $n_+$ and $n_-$ in these two bands (note that $n = n_+ + n_-$). For instance in the inset of Fig. 8(b), $k_{F+}$ and $k_{F-}$ correspond to the intersection of the red line with the two blue lines. This procedure is then repeated so that we obtain self-consistent solutions of $n_+$ and $n_-$ as $U$ (and hence $E_F$) is swept through the valence band, just as is done in experiments. The values of the frequency peaks are related to the carrier densities through re-scaling by a factor of $\frac{h}{2e}$.

Having determined self-consistently the relation between $n_\pm$ and $U$, we proceed to compare theory with experimental results. This is done by plotting the frequency splitting $\delta f$ as a function of the average frequency $\overline{f}$. Here both quantities can be extracted directly and independently from the measured data in Fig. 7(d) (note that $\overline{f}\propto V_g$). In Fig. 8(b), the empty circles correspond to the experimental data measured as $E_F$ is swept through the valence band while the colored lines represent three possible theoretical best-fit plots obtained for different combinations of SOI values $(\lambda_R, \lambda)=(15, 0)$ meV (red solid line), $(\lambda_R, \lambda)=(13, -5)$ meV (green, dashed line), and $(\lambda_R, \lambda)=(13, 5)$ (blue, dashed-dotted line). In this way a very satisfactory fit to the data can be obtained. Upon varying $\lambda_R$ and $\lambda$ over a broad interval, we find that the range of values for which a good agreement is found is $\lambda_R \simeq 10-15$ meV and $\lambda \simeq 0-6$ meV (larger values of $\lambda_R$ constrain $\lambda$ to smaller values to fit the data).

Theory can also reproduce the data obtained when $E_F$ is swept through the conduction band, but in that case the range of carrier density for which a beating is observed experimentally is smaller (see Fig. 7 (d)), and thus only a few data points are present in the $\delta f$-vs-$\overline{f}$ plot (again $\overline{f} \propto V_g$). In the conduction band, the best-fit values of $\lambda_R$ and $\lambda$ are in the range between 5 and 8 meV. Although in this case SOI appears to be slightly weaker than in the valence band, the smaller amount of data makes it more difficult to determine the two parameters accurately.

Our analysis of the beating patterns observed in the SdH measurements therefore indicates that the strong SOI induced in graphene by proximity with TMD is of a band origin. The SOI strength extracted from the comparison between theory and experiments is in the same ballpark, although somewhat larger than the values estimated in Ref.~\cite{Wang2015} from {\it ab-initio} calculations. The similarity of the SOI strength estimated from the data and that found from {\it ab-initio} calculations supports the consistency of our analysis. We therefore conclude that the characteristic magnitude of the interfacially-induced SOI in graphene is about 10 meV. This value is more than a hundred times larger than the SOI intrinsically present in pristine graphene~\cite{Huertas2006,Min2006,Konschuh2010}. This is also in line with the conclusion inferred from the behavior of the spin-relaxation time $\tau_{\rm so}$ obtained from the analysis of the WAL effect above.

\section{Conclusion}

The main conclusion that can be drawn from the measurements presented above is that the interfacially induced SOI is dominated by spin-orbital splitting in the graphene band structure. This generalizes to spin-dependent phenomena the results obtained in graphene-on-hBN moire superlattices, where interfacial interactions alter the graphene band structure by producing secondary Dirac points and creating a gap at the main Dirac point. Our second conclusion is that the interfacially induced SOI is extremely robust. A strong SOI is induced irrespective of the specific TMD material used, of the graphene and TMD lattice alignment angle, of the thickness of the graphene multilayer (which we tested up to three layers), or of the position of the Fermi level in the graphene band.

Most tellingly, we observe a pronounced magneto-conductivity due to weak antilocalization in all devices that we have measured at low temperatures, where the transport regime varied from fully diffusive up to nearly ballistic. None of the devices exhibited a positive magneto-conductivity due to weak-localization. This indicates that SOI in graphene was --in all regimes investigated-- sufficiently strong to cause a full precession of the electron spin even for the shortest trajectories that contribute to electron interference. These observations appear to be only compatible with a band origin of strong SOI, a conclusion that is  confirmed by the experimental observation of beating patterns and a splitting in the frequency of the Shubnikov-de-Hass oscillations in high-quality bilayer devices. The evolution in the magnitude of the splitting that is observed upon varying carrier density indicates that the dominant contribution to the interfacially induced SOI is of the Rashba type. A quantitative analysis of the data indicates that the interaction coupling constant for the Rashba term is as large as 10-15 meV, whereas the strength of the other SOI term that couples spin and valley degrees of freedom is about 5-6 meV or smaller. These values correspond to a remarkably strong SOI, especially in comparison to the minute values of the intrinsic SOI in graphene which are only 20-40 $\mu$eV~\cite{Huertas2006,Min2006,Konschuh2010}.

It is remarkable that such a large interfacial SOI can be induced without causing any damage to the electronic properties of graphene. Our highest quality devices exhibited carrier mobility reaching up to 160,000 cm$^2$/Vs and carrier density inhomogeneity of only $\simeq 2\times10^{10}$ cm$^{-2}$, which is comparable to best graphene-on-hBN devices~\cite{Dean2013,Ponomarenko2013,Hunt2013,Young2012,Taychatanapat2013}. The possibility to achieve such a high quality both in terms of carrier mobility and density homogeneity will be crucial for probing the predicted topologically insulating states that may be realized in graphene-based systems~\cite{Kane2005,Kane20052,Wang2015}.

In that regard we also note that, since Rashba turns out to be the dominant SOI coupling, a gap opening between valence and conduction bands in charge neutral graphene-on-TMD (leading to a topologically insulating state) is not expected to occur~\cite{Wang2015}. However, diverse strategies are available to change the situation. For instance, encapsulating graphene in between two TMD crystals may result in a smaller asymmetry of the device structure, causing a decrease in the intensity of the Rashba term, with other spin-valley SOI contribution becoming stronger. Under these conditions, a topological insulating state may be engineered in graphene with a band gap of several meV~\cite{Wang2015} and, if so, the ability to achieve very high carrier mobility and small density inhomogeneity demonstrated in this work will be essential for probing the occurrence of edge transport in the presence of an insulating bulk.

Another interesting possibility opened up by the results presented here is achieving gate control of SOI in graphene based system. A simple strategy in this regard is to employ dual-gated BLG devices, using a TMD layer as gate insulator on one side  and a hBN layer on the other side. The application of a perpendicular electric field in such a structure will lead to a band gap opening at the charge neutrality point~\cite{Oostinga2008,Weitz2010,Velasco2012}. In this regime, electronic states at the top of the valence band and at the bottom of the conduction band have their wave-functions localized on one of the two BLG layers (depending on the sign of the perpendicular electric field)~\cite{Young20112}. Since  the interfacially induced SOI is present only in the BLG layer which is in direct contact with the TMD, such a configuration will allow switching SOI on and off by tuning the position of the Fermi level and the polarity of the electric field responsible for the gap opening.

Finally, strong interfacially induced SOI in graphene-on-TMDs opens countless opportunities for investigating novel physical phenomena under controlled conditions. For instance, devices can be realized by employing ferromagnetic electrical contacts that enable injection and detection of spins in graphene~\cite{Tombros2007,Han2014}. In these systems the dynamics of spin polarized carriers will be controlled, and altered in an interesting way, by the interfacially induced SOI. These systems will also help to gain new insight into the subtle phenomena originating from spin-Hall effect and inverse spin-Hall effect ~\cite{Jungwirth2012}. These examples illustrate that the ability to engineer the properties of electronic systems in van der Waals heterostructures through a layer-by-layer assembly  --demonstrated here for the case of SOI in graphene-- opens up a wide range of exciting opportunities for realizing and exploring new physical phenomena.

\section{Acknowledgements}
We gratefully acknowledge technical assistance from A. Ferreira. ZW, DK, DM, and AFM also gratefully acknowledge financial support from the Swiss National Science Foundation, the NCCR QSIT, and the EU Graphene Flagship Project. JK is supported by the National Science Scholarship from the Agency for Science, Technology and Research (A*STAR).


\begin{thebibliography}{60}
\bibitem [{sdh()}]{sdhfitting}%
  \BibitemOpen
  \href@noop {} {}\bibinfo {note} {More specifically, for a given $E_F$, we
  compute the total excess (i.e. relative to the background density of positive
  charges) electron density on each layer, including those from the completely
  filled bands. Due to the layer asymmetry of the system, eigenstates from the
  completely filled bands are also partitioned unequally between the two layers
  and thus also contribute (albeit independent of $E_F$), to the inter-layer
  potential $U_n$. $U_n$ is then given by $C_{BLG} U_n = e \delta n_{total,
  2}$, where $C_{BLG}$ is the capacitance per unit area of the bilayer graphene
  (assuming vacuum dielectric constant $\epsilon_0$) and $\delta n_{total, 2}$
  is the total excess electron density on the layer further away from the
  substrate.}\BibitemShut {Stop}%
\end{thebibliography}
%

\end{document}